\documentclass[12pt]{article}

\usepackage{amssymb}
\usepackage{amsmath}
\newtheorem{theorem}{Theorem}{}
\newtheorem{definition}{Definition}{}
\def\func#1{\mathop{\rm #1}}%
\def\limfunc#1{\mathop{\rm #1}}%
\def\Box{\square}

\begin{document}

\title{The ``Bootstrap Program'' for Integrable Quantum Field Theories in 
1+1 Dim
\footnote{To appear in
'FROM INTEGRABLE MODELS TO GAUGE THEORIES',
{\em in Honor of Sergei Matinyan.}
Eds. V.G.Gurzadyan and A.G.Sedrakian
(World Scientific Publications).
Based on talks presentend at the Conference 
%"From QCD to Integrable Models: Old Results and New Developments", 
at Nor Amberd, Yerevan (Armenia), September the 19th to 25th, 2001.}
}

\author{H. Babujian\thanks{%
Permanent address: Yerevan Physics Institute, Alikhanian Brothers 2,
Yerevan, 375036 Armenia.} \thanks{
e-mail: babujian@lx2.yerphi.am, babujian@physik.fu-berlin.de}~ and M.
Karowski\thanks{
e-mail: karowski@physik.fu-berlin.de} \\
Institut f\"ur Theoretische Physik\\
Freie Universit\"at Berlin,\\
Arnimallee 14, 14195 Berlin, Germany}

\maketitle

\begin{abstract}
The purpose of the ``bootstrap program'' is to construct
integrable quantum field theories in 1+1 dimensions in terms of their
Wightman functions explicitly. As an input the integrability and general
assumptions of local quantum field theories are used. The object is to be
achieved in tree steps: 1) The S-matrix is obtained using a qualitative
knowledge of the particle spectrum and the \emph{Yang-Baxter equations}. 2)
Matrix elements of local operators are calculated by means of the ``form
factor program'' using the S-matrix as an input. 3) The Wightman functions
are calculated by taking sums over intermediate states. The first step has
been performed for a large number of models and also the second one for
several models. The third step is unsolved up to now. Here the program is
illustrated in terms of the sine-Gordon model alias the massive Thirring
model. Exploiting the ``off-shell'' Bethe Ansatz we propose general formulae
for form factors. For example the n-particle matrix element for all higher
currents are given and in particular all eigenvalues of the higher conserved
charges are calculated. Furthermore quantum operator equations are obtained
in terms of their matrix elements, in particular the quantum sine-Gordon
field equation. Exact expressions for the finite wave function and mass
renormalization constants are calculated.
\end{abstract}

\section{Introduction}

More than fifty years ago, Heisenberg \cite{Heisen} pointed out the
importance of studying analytic continuations of scattering amplitudes into
the complex momentum plane. The first concrete investigations in this
direction were carried out by Jost \cite{Jost} and Bargmann \cite{Bargmann},
initially for non-relativistic scattering processes. The original ideas
turned out to be very fruitful and lead to interesting results on-shell,
i.e. for the S-matrix \cite{ELOP}, as well as off-shell, that is for the
two-particle form factors \cite{Barton}. Once one restricts ones attention to
1+1 dimensional integrable theories, the n-particle scattering matrix
factories into two particle S-matrices and the approach, now usually
referred to as the 'bootstrap program', reveals its full strength.

The `bootstrap program \cite{K2}. for integrable quantum field theories in
1+1-dimensions' does not start with any classical Lagrangian, but this
program classifies integrable quantum field theoretic models and in addition
it provides their explicit exact solutions in term of all Wightman
functions. We have contact with the classical models only, when at the end
we compare our exact results with Feynman graph expansions which are based
on the Lagrangians.

One of the authors (M.K.) et al. \cite{KTTW}. formulated the on-shell program
i.e. the exact determination of the scattering matrix using the Yang-Baxter
equations. Off-shell quantities, namely form factors were first investigated
by Vergeles and Gryanik \cite{VG} in the sinh-Gordon model and by Weisz \cite
{W} in the sine-Gordon model. The concept of generalized form factors was
introduced by one of the authors (M.K.) et al. \cite{KW}. In this article
consistency equations were formulated which are expected to be satisfied by
these objects. Thereafter this approach was developed further and studied in
the context of several explicit models by Smirnov \cite{Sm} who proposed the
form factor equations $(i)-(v)$ (see below) as extensions of similar
formulae in the original article \cite{KW}. The formulae were proven by the
authors et al. \cite{BFKZ}. There is a large number of more recent papers 
\cite{CM}-\cite{BK2} on form factors. Also there is a nice application \cite
{GNT,CET} of form factors in condensed matter physics. The one dimensional
Mott insulators can be described in terms of the quantum sine-Gordon model.

Finally the Wightman functions are obtained by taking sums and integrals
over intermediate states. The explicit evaluation\footnote{%
Of course one may also adopt a very practical point of view and resort to
the well-known fact that the first terms of the series expansion of
correlation functions in terms of form factors decrease very rapidly.
Consequently correlations functions may be approximated very often quite
well by simply including the low particle number form factor into the
expansion.} of all these integrals and sums remains an open challenge for
almost all theories, except the Ising model \cite{B-J,BKW}. A progress
towards a solution of this problem has recently been achieved by Korepin et
al. \cite{Korepin}.

An entirely different method, the Bethe Ansatz \cite{BA}, was initially
formulated in order to solve the eigenvalue problem for certain integrable
Hamiltonians. The approach has found applications in the context of numerous
models and has led to a detailed study of various mass spectra and
S-matrices. The original techniques have been refined into several
directions, of which in particular the so-called ``off-shell'' Bethe Ansatz,
which was originally formulated by one of the authors (H.B.) \cite{OSBA,BKZ},
will be exploited for our purposes. This version of the Bethe Ansatz paves
the way to extend the approach to the off-shell physics and opens up the
intriguing possibility to merge the two methods, that is the form factor
approach and the Bethe Ansatz. The basis for this opportunity lies in the
observation, that the ``off-shell'' Bethe Ansatz captures the vectorial
structure of Watson's equations. These are matrix difference equations
giving rise to a matrix Riemann-Hilbert problem which is solved by an
''off-shell'' Bethe Ansatz.

\section{The ``bootstrap program''}

The `bootstrap program for integrable quantum field theories in
1+1-dimensions' \emph{classifies} integrable quantum field theoretic models
and in addition it provides their explicit exact solutions in term of all
Wightman functions. The results are obtained in three steps:

\begin{enumerate}
\item  The S-matrix is calculated by means of general properties such as
unitarity and crossing, the Yang-Baxter equations (which are a consequence
of integrability) and the additional assumption of `maximal analyticity'.
This means that the two-particle S-matrix is an analytic function in the
physical plane (of the Mandelstam variable $(p_{1}+p_{2})^{2}$) and
possesses only those poles there which are of physical origin. The only
input which depends on the model is the assumption of a particle spectrum.
Usually it belongs to representations of a symmetry\footnote{%
Typically there is a correspondence of fundamental representations with
multiplets of particles.}.

\item  Generalized form factors which are matrix elements of local operators 
\[
^{out}\left\langle \,p_{m}^{\prime },\ldots ,p_{1}^{\prime }\left| \mathcal{O%
}(x)\right| p_{1},\ldots ,p_{n}\,\right\rangle ^{in}\, 
\]
are calculated by means of the S-matrix. More precisely, the equations $%
(i)-(v)$ given below are solved. These equations follow from LSZ-assumptions
and again the additional assumption of `maximal analyticity'.

\item  The Wightman functions are obtained by inserting a complete set of
intermediate states. In particular the two point function for a hermitian
operator $\mathcal{O}(x)$ reads 
\begin{multline*}
\langle \,0\left| \mathcal{O}(x)\,\mathcal{O}(0)\right| 0\,\rangle
=\sum_{n=0}^{\infty }\frac{1}{n!}\int \dots \int \frac{dp_{1}\ldots dp_{n}}{%
\,(2\pi )^{n}2\omega _{1}\dots 2\omega _{n}} \\
\times \left| \left\langle \,0\left| \mathcal{O}(0)\right| p_{1},\ldots
,p_{n}\,\right\rangle ^{in}\right| ^{2}e^{-ix\sum p_{i}}.
\end{multline*}
Up to now a proof that these sums converge does not exists.
\end{enumerate}

\section{Integrability}

Integrability in (quantum) field theories means that there exits $\infty $%
-many conservation laws 
\[
\partial _{\mu }J_{L}^{\mu }(t,x)=0\quad (L=\pm 1,\pm 3,\dots )\,. 
\]
A consequence of such conservation laws in 1+1 dimensions is that the
n-particle S-matrix is a product of 2-particle S-matrices 
\[
S^{(n)}(p_{1},\dots ,p_{n})=\prod_{i<j}S_{ij}(p_{i},p_{j})\,. 
\]
If there exist backward scattering the 2-particle S-matrices will not
commute and one has to specify the order. In particular for the 3-particle
S-matrix there are two possibilities 
\begin{gather*}
\mathbf{S^{(3)}=S_{12}S_{13}S_{23}=S_{23}S_{13}S_{12}} \\[2mm]
~~~~~~~~~~ 
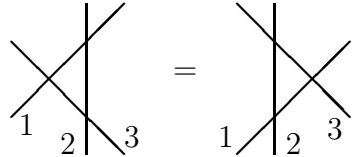
\begin{array}{c}
\unitlength5mm\begin{picture}(9,4) \thicklines \put(0,1){\line(1,1){3}}
\put(0,3){\line(1,-1){3}} \put(2,0){\line(0,1){4}} \put(4.3,2){$=$}
\put(6,0){\line(1,1){3}} \put(6,4){\line(1,-1){3}} \put(7,0){\line(0,1){4}}
\put(.2,.5){$1$} \put(1.3,0){$2$} \put(3,.2){$3$} \put(5.5,.2){$1$}
\put(7.3,0){$2$} \put(8.4,.4){$3$} \end{picture}
\end{array}
\end{gather*}
which yield the \textbf{``Yang-Baxter Equation''.}\\[3pt]
\textbf{Examples }of integrable models in 1+1 - dimensions are the \textbf{%
sine-Gordon} model defined by the classical field equation 
\[
\ddot{\varphi}(t,x)-\varphi ^{\prime \prime }(t,x)+\frac{\alpha }{\beta }%
\sin \beta \varphi (t,x)=0 
\]
and the massive Thirring model defined by the classical Lagrangian 
\[
\mathcal{L}=\bar{\psi}(i\gamma \partial -m)\psi -\tfrac{1}{2}\,g\,\,\bar{\psi%
}\gamma ^{\mu }\psi \,\bar{\psi}\gamma _{\mu }\psi \,. 
\]
Coleman \cite{Co} proved that both models are equivalent on the quantum level.

Further integrable models are: $Z_{N}$-Ising models, nonlinear $\sigma $%
-models, Gross-Neveu models, Toda models etc. In the following most of the
formulae and explicit solutions are given for the sine-Gordon alias massive
Thirring model although often corresponding results exist also for other
models.

\section{The S-matrix}

For the Sine-Gordon alias massive Thiring model the particle spectrum
consists of: soliton, anti-soliton and breathers (as soliton anti-soliton
bound states). Since backward scattering can only appear for particles with
the same mass, the two-particle S-matrix is of the form 
\[
S(\theta )=\left( 
\begin{array}{ccccccc}
u &  &  &  &  &  &  \\ 
& t & r &  &  &  &  \\ 
& r & t &  &  &  &  \\ 
&  &  & u &  &  &  \\ 
&  &  &  & S_{sb} &  &  \\ 
&  &  &  &  & S_{bb} &  \\ 
&  &  &  &  &  & \ddots
\end{array}
\right) 
\]
where the rapidity difference $\theta =|\theta _{1}-\theta _{2}|$ is defined
by\newline
$p_{i}=m_{i}(\cosh \theta _{i},\sinh \theta _{i})$.

We start with the soliton (anti-soliton) S-matrix: 
\[
S_{\alpha \;\beta }^{\beta ^{\prime }\alpha ^{\prime }}= 
\begin{array}{c}
\unitlength2mm\begin{picture}(5,6.3) \thicklines \put(0,1){\line(1,1){4}}
\put(2,3){\makebox(0,0){$\bullet$}} \put(4,1){\line(-1,1){4}}
\put(0,-.5){$\alpha$} \put(3.5,-.5){$\beta$} \put(3.5,5.8){$\alpha'$}
\put(0,5.8){$\beta'$} \end{picture}
\end{array}
:\quad S_{ss}^{ss}=u,\quad S_{s\bar{s}}^{\bar{s}s}=t,\quad S_{s\bar{s}}^{s%
\bar{s}}=r 
\]
$s=$ soliton, $\bar{s}=$ anti-soliton. As input conditions we have:

\begin{enumerate}
\item  {Unitarity} $S(-\theta )S(\theta )=1$ 
\begin{eqnarray*}
u(-\theta )u(\theta ) &=&1 \\
t(-\theta )t(\theta )+r(-\theta )r(\theta ) &=&1 \\
t(-\theta )r(\theta )+r(-\theta )t(\theta ) &=&0
\end{eqnarray*}

\item  {`Crossing'} 
\[
u(i\pi -\theta )=t(\theta )~,~~r(i\pi -\theta )=r(\theta ) 
\]

\item  {Yang-Baxter} 
\[
r(\theta _{12})u(\theta _{13})r(\theta _{23})+t(\theta _{12})r(\theta
_{13})t(\theta _{23})=u(\theta _{12})r(\theta _{13})u(\theta _{23}) 
\]

\item  `Maximal analyticity':

$S(\theta )$ is meromorphic and all poles in the `physical strip' $0\leq 
\func{Im}\theta \leq \pi $ have a physical interpretation; in particular all
bound states correspond to simple poles. An S-matrix satisfying this
condition is also called `minimal'. For couplings $g<0$ (in the language of
the massive Thirring model) there are no soliton anti-soliton bound states.
Therefore in this region of the coupling constant $S(\theta )$ is
holomorphic in the `physical strip' $0\leq \func{Im}\theta \leq \pi $.
\end{enumerate}

The S-matrix bootstrap using the {Yang-Baxter} relations was proposed by
Karowski, Thun, Truong and Weisz \cite{KTTW}. It was shown in this article
that the 'minimal' {general solution} of these equations is 
\[
u(\theta ,\nu )=\exp \int_{0}^{\infty }\frac{dt}{t}\,\frac{\sinh \frac{t}{2}%
(1-\nu )}{\sinh \frac{\nu t}{2}\,\cosh \frac{t}{2}}\,\sinh t\frac{\theta }{%
i\pi }\,. 
\]
This S-matrix was first obtained by Zamolodchikov \cite{Za} from the
extrapolation of semi-classical expressions. It has been checked in
perturbation theory. The free parameter $\nu $ is related to the coupling
constants by 
\[
\frac{1}{\nu }=\frac{8\pi }{\beta ^{2}}-1=1+\frac{2g}{\pi }\,. 
\]
The second equation is due to Coleman and the first one is obtained by
analyzing the pole structure of the amplitude $u(\theta ,\nu )$. The
assumption of 'maximal analyticity' and comparison with the known
semi-classical bound state spectrum implies the identification of the
parameter $\nu $ (see below).

\subsection*{Bound states}

Let $\gamma $ be a bound state of the particles $\alpha $ and $\beta $ with
mass 
\[
m_{\gamma }=\sqrt{m_{\alpha }^{2}+m_{\beta }^{2}+2m_{\alpha }m_{\beta }\cos a%
}~,~~~(0<a<\pi ) 
\]
then the 2 particle S-matrix has a pole such that 
\begin{eqnarray*}
i\limfunc{Res}_{\theta =ia}\dot{S}_{\alpha \beta }^{\beta ^{\prime }\alpha
^{\prime }}(\theta ) &=&\Gamma _{\gamma }^{\beta ^{\prime }\alpha ^{\prime
}}\,\Gamma _{\alpha \beta }^{\gamma } \\[5mm]
i\limfunc{Res} 
\begin{array}{c}
\unitlength2.7mm\begin{picture}(5,6) \thicklines \put(0,1){\line(1,1){4}}
\put(2,3){\makebox(0,0){$\bullet$}} \put(4,1){\line(-1,1){4}}
\put(0,-.3){$\alpha$} \put(3.5,-.3){$\beta$} \put(3.5,5.5){$\alpha'$}
\put(0,5.5){$\beta'$} \end{picture}
\end{array}
~~ &=&~~~ 
\begin{array}{c}
\unitlength2mm\begin{picture}(5,8) \thicklines \put(0,1){\line(1,1){2}}
\put(2,3){\makebox(0,0){$\bullet$}} \put(4,1){\line(-1,1){2}}
\put(2,3){\line(0,1){2}} \put(2,5){\makebox(0,0){$\bullet$}}
\put(2,5){\line(1,1){2}} \put(2,5){\line(-1,1){2}} \put(0,-.5){$\alpha$}
\put(3.5,-.5){$\beta$} \put(2.5,3.7){$\gamma$} \put(3.5,7.7){$\alpha'$}
\put(0,7.7){$\beta'$} \end{picture}
\end{array}
\end{eqnarray*}
where $a$ is called the fusion angle and $\Gamma _{\alpha \beta }^{\gamma }$
is the `bound state intertwiner' \cite{K1,BK}. The \textbf{bound state
S-matrix, }that is the scattering matrix of the bound state (12) with a
particle 3, is obtained by the ``bootstrap equation'' \cite{K1} 
\begin{eqnarray*}
S_{(12)3}(\theta _{(12)3})\,\Gamma _{12}^{(12)} &=&\Gamma
_{12}^{(12)}\,S_{13}(\theta _{13})S_{23}(\theta _{23})\, \\[5mm]
\begin{array}{c}
\unitlength3mm\begin{picture}(5,5) \thicklines\put(2,2){\line(1,1){2}}
\put(4.5,1.5){\line(-1,1){3}} \put(2,2){\line(-4,-1){2}}
\put(2,0){\line(0,1){2}} \put(0,.4){$1$} \put(1,0){$2$} \put(3.5,.7){$3$}
\put(4.2,4){$(12)$} \put(2,2){\makebox(0,0){$\bullet$}} \end{picture}
\end{array}
~~~ &=&~~~ 
\begin{array}{c}
\unitlength3mm\begin{picture}(5,5) \thicklines \put(3,3){\line(1,1){1.2}}
\put(4,0){\line(-1,1){4}} \put(0,2){\line(3,1){3}} \put(2,0){\line(1,3){1}}
\put(0,.8){$1$} \put(1,0){$2$} \put(4,.4){$3$} \put(4.4,4){$(12)$}
\put(3,3){\makebox(0,0){$\bullet$}} \end{picture}
\end{array}
\end{eqnarray*}
where we use the usual short notation of matrices acting in the spaces
corresponding to the particles 1, 2, 3 and (12).

As an example we consider the sine-Gordon alias massive Thirring model. For $%
\nu >1$ (i.e. $g<0$) there are no soliton anti-soliton bound states and the
S-matrix $S(\theta )$ is analytic in the physical strip $0\leq \func{Im}%
\theta \leq \pi $. For $\nu <1$ there are poles of the S-matrix element $S_{s%
\bar{s}}(\theta )$ at $\theta =i\pi (1-k\nu )$ for $k=1,\dots <1/\nu $ which
correspond to bound states with masses $m_{k}=2M\sin \tfrac{\pi }{2}\nu k$
where $M$ is the soliton mass. This mass spectrum coincides with the
semiclassical \cite{DHNKF} breather spectrum if the parameter $\nu $ is
related to the sine-Gordon coupling constant $\beta $ as noted above. The
pole at $\theta =i\pi (1-\nu )$ correspond in particular to 
\[
\text{soliton + anti-soliton}\longrightarrow \text{lowest breather.} 
\]
The ``bootstrap equations'' yield the breather-soliton S-matrix \cite{KT} 
\[
S_{bs}(\theta )=t(\theta +\tfrac{1}{2}i\pi (1-\nu ))\,u(\theta -\tfrac{1}{2}%
i\pi (1-\nu ))=\frac{\sinh \theta +i\sin \tfrac{1}{2}\pi (1+\nu )}{\sinh
\theta -i\sin \tfrac{1}{2}\pi (1+\nu )} 
\]
and in a second step the 2-breather S-matrix \cite{KT} 
\[
S_{bb}(\theta )=S_{bs}(\theta +\tfrac{1}{2}i\pi (1-\nu ))\,S_{bs}(\theta -%
\tfrac{1}{2}i\pi (1-\nu ))=\frac{\sinh \theta +i\sin \pi \nu }{\sinh \theta
-i\sin \pi \nu } 
\]

\section{Form factors}

\begin{definition}
For a local operator $\mathcal{O}(x)$ the generalized form factors \cite{KW}
are defined as 
\[
\,\mathcal{O}_{\alpha _{1}\dots \alpha _{n}}\left( \theta _{1},\dots ,\theta
_{n}\right) =\langle \,0\,|\,\mathcal{O}(0)\,|\,p_{1},\dots ,p_{n}\,\rangle
_{\alpha _{1}\dots \alpha _{n}}^{in} 
\]
for $\theta _{1}>\dots >\theta _{n}$ (for other orders of the rapidities
they are defined by analytic continuation). The index $\alpha _{i}$%
\thinspace denotes the type of the particle with momentum $p_{i}$. We also
use the short notations $\mathcal{O}_{\underline{\alpha }}(\underline{\theta 
})$ or $\mathcal{O}_{1\dots n}(\underline{\theta })$.
\end{definition}

For the sine Gordon model $\alpha =$ soliton, anti-soliton or breathers.
First we restrict ourselves to solitonic states. Similar as for the S-matrix
`maximal analyticity' for generalized form factors means again that they are
meromorphic and all poles in the `physical strips' $0\leq \func{Im}\theta
_{i}\leq \pi $ have a physical interpretation. Together with the usual
LSZ-assumptions \cite{LSZ} of local quantum field theory the form factor
equations can be derived

\begin{enumerate}
\item[$(i)$]  Watson's equations: 
\[
\mathcal{O}_{\dots ij\dots }(\dots ,\theta _{i},\theta _{j},\dots )=\mathcal{%
O}_{\dots ji\dots }(\dots ,\theta _{j},\theta _{i},\dots )\,S_{ij}(\theta
_{i}-\theta _{j}) 
\]

\item[$(ii)$]  Crossing relations: 
\begin{eqnarray*}
_{\bar{\alpha}_{1}}\langle \,p_{1}\,|\,\mathcal{O}(0)\,|\,p_{2},\dots
,p_{n}\,\rangle _{\alpha _{2}\dots \alpha _{n}}^{in,conn.} &=&\mathcal{O}%
_{\alpha _{1}\alpha _{2}\dots \alpha _{n}}(\theta _{1}+i\pi ,\theta
_{2},\dots ,\theta _{n}) \\
&=&\mathcal{O}_{\alpha _{2}\dots \alpha _{n}\alpha _{1}}(\theta _{2},\dots
,\theta _{n},\theta _{1}-i\pi )
\end{eqnarray*}

\item[$(iii)$]  Recursion relations: 
\[
\limfunc{Res}\limits_{\theta _{12}=i\pi }\mathcal{O}_{1\dots n}(\theta
_{1},\dots )=2i\,\mathbf{C}_{12}\,\,\mathcal{O}_{3\dots n}(\theta _{3},\dots
)\left( \mathbf{1}-S_{2n}\dots S_{23}\right) 
\]
where $\mathbf{C}_{12}$ is the charge conjugation matrix

\item[$(iv)$]  Bound state form factors equation: 
\[
\limfunc{Res}\limits_{\theta _{12}=ia}\mathcal{O}_{123\dots n}(\underline{%
\theta })=\mathcal{O}_{(12)3\dots n}(\theta _{(12)},\underline{\theta }%
^{\prime })\,\sqrt{2}\,\Gamma _{12}^{(12)} 
\]
where $a$ is the fusion angle.

\item[$(v)$]  Lorentz invariance: 
\[
\mathcal{O}_{1\dots n}(\theta _{1}+u,\dots ,\theta _{n}+u)=e^{su}\,\mathcal{O%
}_{1\dots n}(\theta _{1},\dots ,\theta _{n}) 
\]
where $s$ is the ``spin'' of $\mathcal{O}$.
\end{enumerate}

These equations have been proposed by Smirnov \cite{Sm} as generalizations of
equations derived in the original articles \cite{KW,BKW,K2}. They have been
proven \cite{BFKZ} by means of LSZ-assumptions and `maximal analyticity'.
They hold in this form for bosons; for fermions or more generally for anyons
there are some additional phase factors. If we write the form factors as a
formal sum of Feynman graphs and use the additional rule that a line
changing the 'time' direction also changes a particle to an anti-particle
and changes the rapidity $\theta \rightarrow \theta \pm i\pi $ we may
depicted the equations $(i)-(iv)$ as in figure \ref{f0}. 
\begin{figure}[tbh]
\[
\begin{array}{rrcl}
(i) & 
\begin{array}{c}
\unitlength3.2mm\begin{picture}(7,3) \put(3.5,2){\oval(7,2)}
\put(3.5,2){\makebox(0,0){${\cal O}$}} \put(1,0){\line(0,1){1}}
\put(3,0){\line(0,1){1}} \put(4,0){\line(0,1){1}} \put(6,0){\line(0,1){1}}
\put(1.4,.5){$\dots$} \put(4.4,.5){$\dots$} \end{picture}
\end{array}
~~ & = & ~~ 
\begin{array}{c}
\unitlength3.2mm\begin{picture}(7,4) \put(3.5,3){\oval(7,2)}
\put(3.5,3){\makebox(0,0){${\cal O}$}} \put(1,0){\line(0,1){2}}
\put(3,0){\line(1,2){1}} \put(4,0){\line(-1,2){1}} \put(6,0){\line(0,1){2}}
\put(1.4,1){$\dots$} \put(4.4,1){$\dots$} \end{picture}
\end{array}
\\ 
(ii) & 
\begin{array}{c}
\unitlength3.2mm\begin{picture}(6,4) \put(3,2){\oval(6,2)[]}
\put(3,2){\makebox(0,0){${\cal O}_{conn.}$}} \put(1,0){\line(0,1){1}}
\put(2.4,0.5){$\dots$} \put(5,0){\line(0,1){1}} \put(3,3){\line(0,1){1}}
\end{picture}
\end{array}
~~ & = & ~~ 
\begin{array}{c}
\unitlength3.2mm\begin{picture}(6,4) \put(1,1){\oval(2,2)[b]}
\put(3.5,2){\oval(5,2)} \put(3.5,2){\makebox(0,0){${\cal O}$}}
\put(3,0){\line(0,1){1}} \put(5,0){\line(0,1){1}} \put(0,1){\line(0,1){3}}
\put(3.4,.5){$\dots$} \end{picture}
\end{array}
~=~ 
\begin{array}{c}
\unitlength3.2mm\begin{picture}(6,4) \put(5,1){\oval(2,2)[b]}
\put(2.5,2){\oval(5,2)} \put(2.5,2){\makebox(0,0){${\cal O}$}}
\put(1,0){\line(0,1){1}} \put(3,0){\line(0,1){1}} \put(6,1){\line(0,1){3}}
\put(1.4,.5){$\dots$} \end{picture}
\end{array}
\\ 
(iii) & ~~\dfrac{1}{2i}\,\limfunc{Res}\limits_{\theta _{12}=i\pi }~~ 
\begin{array}{c}
\unitlength3.2mm\begin{picture}(6,4) \put(3,2){\oval(6,2)}
\put(3,2){\makebox(0,0){${\cal O}$}} \put(1,0){\line(0,1){1}}
\put(2,0){\line(0,1){1}} \put(3,0){\line(0,1){1}} \put(5,0){\line(0,1){1}}
\put(3.4,.5){$\dots$} \end{picture}
\end{array}
~~ & = & ~~ 
\begin{array}{c}
\unitlength3.2mm\begin{picture}(5,4) \put(.5,0){\oval(1,2)[t]}
\put(3,2){\oval(4,2)} \put(3,2){\makebox(0,0){${\cal O}$}}
\put(2,0){\line(0,1){1}} \put(4,0){\line(0,1){1}} \put(2.4,.5){$\dots$}
\end{picture}
\end{array}
- 
\begin{array}{c}
\unitlength3.2mm\begin{picture}(6,5.5) \put(0,0){\line(0,1){3}}
\put(3,3){\oval(6,4)[t]} \put(3,3){\oval(6,4)[br]} \put(3,0){\oval(4,2)[tl]}
\put(3,3){\oval(4,2)} \put(3,3){\makebox(0,0){${\cal O}$}}
\put(2,0){\line(0,1){2}} \put(4,0){\line(0,1){2}} \put(2.4,1.5){$\dots$}
\end{picture}
\end{array}
\\ 
(iv) & \dfrac{1}{\sqrt{2}}\,\limfunc{Res}\limits_{\theta _{12}=ia} 
\begin{array}{c}
\unitlength3.2mm\begin{picture}(5,3) \put(2.5,2){\oval(5,2)}
\put(2.5,2){\makebox(0,0){${\cal O}$}} \put(1,0){\line(0,1){1}}
\put(2,0){\line(0,1){1}} \put(4,0){\line(0,1){1}} \put(2.4,.5){$\dots$}
\end{picture}
\end{array}
~~ & = & ~~ 
\begin{array}{c}
\unitlength3.2mm%
\begin{picture}(5,4) \put(2.5,3){\oval(5,2)} \put(2.5,3){\makebox(0,0){${\cal O}$}}
\put(1.5,0){\oval(1,2)[t]}
\put(1.5,1){\line(0,1){1}} \put(4,0){\line(0,1){2}} \put(2.4,1){$\dots$} \end{picture}
\end{array}
\end{array}
\]
\caption{\textit{The form factor equations. }}
\label{f0}
\end{figure}

\subsection*{Locality}

It has been proven \cite{Q}\footnote{%
For the case of no bound states this was proven before by Smirnov \cite{Sm}
and Lashkevich \cite{Las}}. that the properties $(i)-(iv)$ of the form
factors together with a general crossing formula \cite{BK} imply locality in
the form of 
\[
^{in}\langle \,\phi \,|\,\left[ \mathcal{O}(x),\mathcal{O}(y)\right]
\,|\,\psi \,\rangle ^{in}=0 
\]
for all matrix elements, if $x-y$ is space like. In the proof one assumes
the convergence of the sum over all intermediate states.

\subsection*{Two-particle form factors}

For the two-particle form factors the form factor equations are easily
understood. The usual assumptions of local quantum field theory yield 
\[
\langle \,0\,|\,\mathcal{O}(0)\,|\,p_{1},p_{2}\rangle ^{in/out}=F\left(
(p_{1}+p_{2})^{2}\pm i\varepsilon \right) =\,F\left( \pm \theta _{12}\right) 
\]
where the rapidity difference is defined by $p_{1}p_{2}=m^{2}\cosh \theta
_{12}$. For integrable theories one has particle number conservation which
implies (for any eigenstate of the two-particle S-matrix) 
\[
\langle \,0\,|\,\mathcal{O}(0)\,|\,p_{1},p_{2}\rangle ^{in}=\langle \,0\,|\,%
\mathcal{O}(0)\,|\,p_{2},p_{1}\rangle ^{out}\,S\left( \theta _{12}\right)
\,. 
\]
Crossing means 
\[
\langle \,p_{1}\,|\,\mathcal{O}(0)\,|\,p_{2}\rangle =F\left( i\pi -\theta
_{12}\right) 
\]
where for one-particle states in- and out-states coincide. Therefore
Watson's equations follow 
\[
\begin{array}{l}
F\left( \theta \right) =F\left( -\theta \right) S\left( \theta \right) 
\vspace{5pt} \\ 
F\left( i\pi -\theta \right) =F\left( i\pi +\theta \right) \,.
\end{array}
\]
For general theories Watson's \cite{Wa} equations only hold below particle
production thresholds. However, for integrable theories there is no particle
production and therefore they hold for all complex values of $\theta $. It
has been shown \cite{KW} that these equations together with ``maximal
analyticity'' have a unique solution. As an example we write the sine-Gordon
(alias massive Thirring model) two-soliton form factor \cite{KW} 
\[
F(i\pi -\theta )=\cosh \tfrac{1}{2}\theta \,\exp \int_{0}^{\infty }\frac{dt}{%
t}\,\frac{\sinh \frac{t}{2}(1-\nu )}{\sinh \frac{\nu t}{2}\,\cosh \frac{t}{2}%
\sinh t}\,\sin ^{2}t\frac{\theta }{2\pi }\,. 
\]

\subsection*{A formula for generalized form factors}

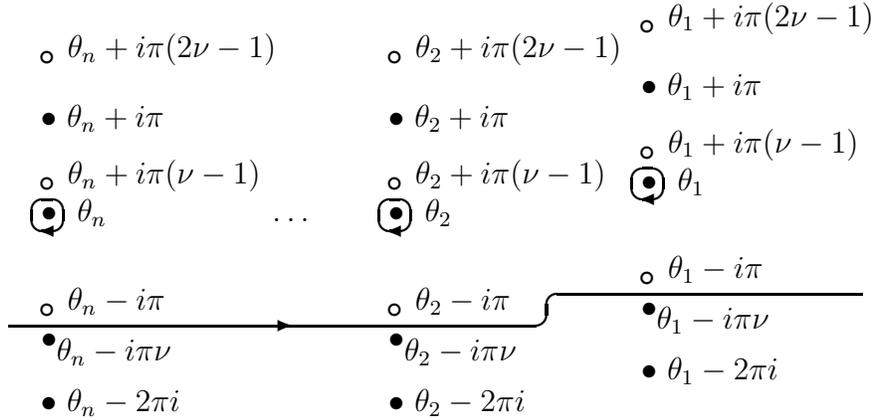
\begin{figure}[tbh]
\[
\unitlength4.2mm%
\begin{picture}(27,13)
\thicklines
\put(1,0){
\put(0,0){$\bullet~\theta_n-2\pi i$}
\put(0,2){$\bullet$}\put(.5,1.6){$\theta_n-i\pi\nu$}
\put(.19,3.2){\circle{.3}~$\theta_n-i\pi$}
\put(0,6){$\bullet~~\theta_n$}
\put(.2,6.2){\oval(1,1)}\put(-.1,5.71){\vector(-1,0){0}}
\put(.19,7.2){\circle{.3}~$\theta_n+i\pi(\nu-1)$}
\put(0,9){$\bullet~\theta_n+i\pi$}
\put(.19,11.2){\circle{.3}~$\theta_n+i\pi(2\nu-1)$}
}
\put(8,6){\dots}
\put(12,0){
\put(0,0){$\bullet~\theta_2-2\pi i$}
\put(0,2){$\bullet$}\put(.5,1.6){$\theta_2-i\pi\nu$}
\put(.19,3.2){\circle{.3}~$\theta_2-i\pi$}
\put(0,6){$\bullet~~\theta_2$}
\put(.2,6.2){\oval(1,1)}\put(-.1,5.71){\vector(-1,0){0}}
\put(.19,7.2){\circle{.3}~$\theta_2+i\pi(\nu-1)$}
\put(0,9){$\bullet~\theta_2+i\pi$}
\put(.19,11.2){\circle{.3}~$\theta_2+i\pi(2\nu-1)$}
}
\put(20,1){
\put(0,0){$\bullet~\theta_1-2\pi i$}
\put(0,2){$\bullet$}\put(.5,1.6){$\theta_1-i\pi\nu$}
\put(.19,3.2){\circle{.3}~$\theta_1-i\pi$}
\put(0,6){$\bullet~~\theta_1$}
\put(.2,6.2){\oval(1,1)}\put(-.1,5.71){\vector(-1,0){0}}
\put(.19,7.2){\circle{.3}~$\theta_1+i\pi(\nu-1)$}
\put(0,9){$\bullet~\theta_1+i\pi$}
\put(.19,11.2){\circle{.3}~$\theta_1+i\pi(2\nu-1)$}
}
\put(9,2.7){\vector(1,0){0}}
\put(0,3.2){\oval(34,1)[br]}
\put(27,3.2){\oval(20,1)[tl]}
\end{picture}
\]
\caption{\textit{The integration contour $\mathcal{C_{\protect\underline{%
\theta }}}$ (for the repulsive case $\nu >1$). The bullets belong to poles
of the integrand resulting from $u(\theta _{i}-u_{j})\,\phi (\theta
_{i}-u_{j})$ and the small open circles belong to poles originating from $%
t(\theta _{i}-u_{j})$ and $r(\theta _{i}-u_{j})$. }}
\label{f}
\end{figure}

We are looking for solutions of the form factor equations $(i)-(v)$. For
generalized form factors for arbitrary numbers of solitons and anti-solitons
we make the Ansatz\footnote{%
This formula using the so called `off-shell Bethe Ansatz' has been
introduced by the authors et al. \cite{BFKZ}. Similar integral representation
have been also introduced by Smirnov \cite{Sm}.} 
\begin{equation}
\fbox{$\rule{0in}{0.17in}~\mathcal{O}_{\underline{\alpha }}(\underline{%
\theta })=\int_{\mathcal{C}_{\underline{\theta }}}dz_{1}\cdots \int_{%
\mathcal{C}_{\underline{\theta }}}dz_{m}\,h(\underline{\theta },{\underline{z%
}})\,p_{n}^{\mathcal{O}}(\underline{\theta },{\underline{z}})\,\Psi _{%
\underline{\alpha }}(\underline{\theta },{\underline{z}})~$}  \label{1}
\end{equation}
which transforms the equations $(i)-(v)$ for the co-vector valued function $%
\mathcal{O}_{\underline{\alpha }}(\underline{\theta })$ into simple
equations $(i^{\prime })-(v^{\prime })$ for scalar functions $p_{n}^{%
\mathcal{O}}(\underline{\theta },{\underline{z})}$ where $n$ is the number
of particles. The later equations are easily solved (see below). To capture
the vectorial structure we use the ``off-shell'' Bethe Ansatz state $\Psi _{%
\underline{\alpha }}(\underline{\theta },{\underline{z}})$ (see below). For
all integration variables $z_{j}$ $(j=1,\dots ,m)$ the integration contours $%
\mathcal{C}_{\underline{\theta }}$ consists of several pieces (see figure~%
\ref{f}). 
The number of integrations $m$ depends on the charge of the operator $q=n-2m$%
. The scalar function $h(\underline{\theta },{\underline{z}})$ is uniquely
determined by the S-matrix 
\[
h(\underline{\theta },{\underline{z}})=\prod_{1\le i<j\le n}F(\theta
_{ij})\prod_{i=1}^{n}\prod_{j=1}^{m}\phi (\theta _{i}-z_{j})\prod_{1\le
i<j\le m}\tau (z_{i}-z_{j})\,, 
\]
with 
\[
\phi (z)=\frac{1}{F(z)\,F(z+i\pi )}~,~~~~\tau (z)=\frac{1}{\phi (z)\,\phi
(-z)}\varpropto \sinh z\sinh z/\nu 
\]
where $F(\theta )$ is the soliton-soliton form factor above. The dependence
of the operator $\mathcal{O}(x)$ enters only through the scalar p-functions $%
p_{n}^{\mathcal{O}}(\underline{\theta },{\underline{z})}$ (see below). We
consider p-functions which satisfy the following conditions:

\begin{itemize}
\item[$(i^{\prime })$]  $p_{n}^{\mathcal{O}}(\underline{\theta },\underline{z%
})$ is symmetric with respect to the $\theta $'s and the $z$'s.

\item[$(ii^{\prime })$]  $p_{n}^{\mathcal{O}}(\underline{\theta },\underline{%
z})=p_{n}^{\mathcal{O}}(\dots ,\theta _{i}-2\pi i,\dots ,\underline{z})$ and
it is a polynomial in $e^{\pm z_{j}}~(j=1,\dots ,m)$.

\item[$(iii^{\prime })$]  $\left\{ 
\begin{array}{l}
p_{n}^{\mathcal{O}}(\theta _{1}=\theta _{n}+i\pi ,\tilde{\underline{\theta }}%
,\theta _{n};\tilde{\underline{z}},z_{m}=\theta _{n}{)}=\dfrac{\varkappa }{m}%
\,p_{n-2}^{\mathcal{O}}(\tilde{\underline{\theta }},\tilde{\underline{z}})+%
\tilde{p}^{(1)}({\underline{\theta }})\vspace{5pt} \\ 
p_{n}^{\mathcal{O}}(\theta _{1}=\theta _{n}+i\pi ,\tilde{\underline{\theta }}%
,\theta _{n};\tilde{\underline{z}},z_{m}=\theta _{1}{)}=\dfrac{\varkappa }{m}%
\,p_{n-2}^{\mathcal{O}}(\tilde{\underline{\theta }},\tilde{\underline{z}})+%
\tilde{p}^{(2)}({\underline{\theta }})
\end{array}
\right. $

where $\tilde{\underline{\theta }}=(\theta _{2},\dots ,\theta _{n-1}),\;%
\tilde{\underline{z}}=(z_{1},\dots z_{m-1})$ and \cite{BFKZ} $\varkappa
=-\left( F^{\prime }(0)\right) ^{2}/\pi $. The functions $\tilde{p}^{(1,2)}({%
\underline{\theta }})$ are non-vanishing only for charge less operators and
they are independent of the $z$s.

\item[$(iv^{\prime })$]  The bound state p-functions are investigated below.

\item[$(v^{\prime })$]  $p_{n}^{\mathcal{O}}(\underline{\theta }+\mu ,%
\underline{z}+\mu )=e^{s\mu }p_{n}^{\mathcal{O}}(\underline{\theta },%
\underline{z})$ where $s$ is the `spin' of the operator $\mathcal{O}(x)$.
\end{itemize}

Again these equations hold in this form for bosons; for fermions or more
generally for anyons there are some additional phase factors.

\begin{theorem}
Let generalized form factors be given by the Ansatz (\ref{1}). They satisfy
the form factor equations $(i)-(v)$ if the p-functions $p_{n}^{\mathcal{O}}(%
\underline{\theta },\underline{z})$ satisfies the equations $(i^{\prime
})-(v^{\prime })$.
\end{theorem}

This theorem has bee proven for odd \cite{BFKZ} and for even \cite{BK} number
of solitonic particles.

\subsection*{The ``off-shell Bethe Ansatz'' state}

As usual one defines the `monodromy matrix' 
\begin{eqnarray*}
T_{1\dots n,0}({\underline{\theta }},\theta _{0}) &=&S_{10}(\theta
_{1}-\theta _{0})\,S_{20}(\theta _{2}-\theta _{0})\cdots S_{n0}(\theta
_{n}-\theta _{0}) \\
&=& 
\begin{array}{c}
\unitlength2.7mm\begin{picture}(10,4)\thicklines \put(0,2){\line(1,0){10}}
\put(2,0){\line(0,1){4}} \put(4,0){\line(0,1){4}} \put(8,0){\line(0,1){4}}
\put(1,0){$1$} \put(3,0){$ 2$} \put(7,0){$ n$} \put(9,.8){$ 0$}
\put(5,1){$\dots$} \end{picture}
\end{array}
\end{eqnarray*}
acting in the tensor product of the `quantum space' and the `auxiliary
space' $V^{1\dots n}\otimes V_{0}$ with $V^{1\dots n}=V_{1}\otimes \cdots
\otimes V_{n}$ (for the sine-Gordon model which correspond to the quantum
group $sl_{q}(2)$ all $V_{i}\cong \Bbb{C}^{2}$). Further one defines as
usual the sub-matrices $A,B,C,D$ by 
\[
T_{1\dots n,0}({\underline{\theta }},z)\equiv \left( 
\begin{array}{cc}
A_{1\dots n}({\underline{\theta }},z) & B_{1\dots n}({\underline{\theta }},z)
\\ 
C_{1\dots n}({\underline{\theta }},z) & D_{1\dots n}({\underline{\theta }},z)
\end{array}
\right) ~. 
\]
The ``pseudo-vacuum'' consists only of solitons 
\[
\Omega _{1\dots n}=s\otimes \cdots \otimes s\,. 
\]
A Bethe Ansatz co-vector in $V_{1\dots n}$ is defined by 
\[
\begin{array}{rcl}
\Psi _{1\dots n}({\underline{\theta }},\underline{z}) & = & \Omega _{1\dots
n}C_{1\dots n}({\underline{\theta }},z_{1})\cdots C_{1\dots n}({\underline{%
\theta }},z_{m})\,. \\ 
\begin{array}{c}
\unitlength3.4mm\begin{picture}(6,4) \thicklines\put(3,2){\oval(6,2)}
\put(3,2){\makebox(0,0){$\Psi$}} \put(1,0){\line(0,1){1}}
\put(5,0){\line(0,1){1}} \put(-.3,0){$\theta_1$} \put(5.3,0){$\theta_n$}
\put(2.5,.5){$\dots$} \end{picture}
\end{array}
~~ & = & ~~~~ 
\begin{array}{c}
\unitlength3.4mm\begin{picture}(7,5.5) \thicklines \put(0,1){\line(1,0){7}}
\put(0,3){\line(1,0){7}} \put(1,0){\line(0,1){4}} \put(5,0){\line(0,1){4}}
\put(.8,4.3){$s$} \put(4.8,4.3){$s$} \put(-.8,2.8){$s$} \put(-.8,.8){$s$}
\put(7.3,2.8){$\bar s$} \put(7.3,.8){$\bar s$} \put(0,-.3){$\theta_1$}
\put(3.8,-.3){$\theta_n$} \put(6,.2){$z_m$} \put(6,3.4){$z_1$}
\put(2.5,2){$\dots$} \put(.3,1.7){$\vdots$} \put(5.3,1.7){$\vdots$}
\end{picture}
\end{array}
\,
\end{array}
\]
The conventional \cite{BA} application of the Bethe Ansatz is to solve an
eigenvalue problem of a spin chain Hamiltonian or of a transfer matrix. The
eigenstates are then given by a discrete set of Bethe Ansatz vectors where
the parameters $z_{i}$ have to satisfy the Bethe Ansatz equations. In the
``off-shell Bethe Ansatz'' \cite{OSBA} the parameter $z_{i}$ are summed or
integrated over as above in our Ansatz for the form factors. For other
models which correspond to groups or quantum groups of higher rank one has
to apply a nested off-shell Bethe Ansatz \cite{BKZ}.

\subsection*{Examples of p-functions}

Form factors for arbitrary numbers of solitons and anti-solitons are given
by the Ansatz above which involves the p-functions. For the sine-Gordon
alias massive Thiring-model there have been proposed the p-functions for
several local operators. \\[3pt]
\textbf{Examples} of operators and their p-functions (up to normalizations)
\cite{BK} 
\[
\mathcal{O}(x)\leftrightarrow p_{n}^{\mathcal{O}}(\underline{\theta },%
\underline{z}) 
\]
{The fundamental fermi field} (charge $q=n-2m=1$) \cite{BFKZ} 
\[
\psi (x)\leftrightarrow \exp \pm \left( \sum_{j=1}^{m}z_{j}-\tfrac{1}{2}%
\sum_{i=1}^{n}\theta _{i}\right) . 
\]
{The fundamental breather field} (charge $q=n-2m=0$) \cite{BK} 
\[
\varphi (x)\leftrightarrow \frac{1}{\sum e^{\theta }\sum e^{-\theta }}\left(
\sum e^{-\theta _{i}}\sum e^{z_{j}}+\sum e^{\theta _{i}}\sum
e^{-z_{j}}\right) . 
\]
{The current} $j^{\mu }=\overline{\psi }\gamma ^{\mu }\psi (x)$ \cite{BK} 
\[
\overline{\psi }\gamma ^{\pm }\psi (x)\leftrightarrow \frac{\pm 1}{\sum
e^{\mp \theta _{i}}}\left( \sum e^{-\theta _{i}}\sum e^{z_{j}}+\sum
e^{\theta _{i}}\sum e^{-z_{j}}\right) . 
\]
{The energy momentum tensor} (with $\rho ,\sigma =+,-$) \cite{BK} 
\[
T{^{\rho \sigma }(x)}\leftrightarrow \rho \frac{\sum e^{\rho \theta _{i}}}{%
\sum e^{-\sigma \theta _{i}}}\left( \sum e^{-\theta _{i}}\sum e^{z_{j}}-\sum
e^{\theta _{i}}\sum e^{-z_{j}}\right) . 
\]
The $\infty $-many conserved currents \cite{BK} 
\[
J_{L}^{\pm }\leftrightarrow \sum_{i=1}^{n}e^{\pm \theta
_{i}}\sum_{j=1}^{m}e^{Lz_{j}}\quad (L=\pm 1,\pm 3,\dots )\,. 
\]

It is easy to check that these p-functions satisfy the equations $(i^{\prime
})-(v^{\prime })$ consistently with the quantum numbers of the operators.

\subsection*{Identification of the operators}

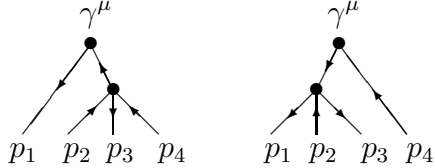
\begin{figure}[h]
\[
\unitlength3mm%
\begin{picture}(19,7)
\put(3,1){\line(1,1){2}} \put(4,2){\vector(1,1){.3}}
\put(5,1){\line(0,1){2}} \put(5,2){\vector(0,-1){.3}}
\put(5,3){\line(-1,2){1}} \put(5,3){\vector(-1,2){.6}}
\put(7,1){\line(-1,1){2}} \put(6,2){\vector(-1,1){.3}}
\put(5,3){\makebox(0,0){$\bullet$}}
\put(4,5){\makebox(0,0){$\bullet$}} \put(4,5){\line(-3,-4){3}}
\put(4,5){\vector(-3,-4){1.5}} \put(3.5,6){$\gamma^\mu$}
\put(.4,0){$p_1$} \put(2.8,0){$p_2$}% \put(3.4,-2){$(a)$}
\put(4.7,0){$p_3$} \put(7,0){$p_4$} \put(12,1){\line(1,1){2}}
\put(13,2){\vector(-1,-1){.3}} \put(14,1){\line(0,1){2}}
\put(14,2){\vector(0,1){.3}} \put(16,1){\line(-1,1){2}}
\put(15,2){\vector(1,-1){.3}} \put(14,3){\makebox(0,0){$\bullet$}}
\put(15,5){\makebox(0,0){$\bullet$}} \put(15,5){\line(-1,-2){1}}
\put(15,5){\vector(-1,-2){.6}} \put(18,1){\line(-3,4){3}}
\put(18,1){\vector(-3,4){1.5}} \put(14.5,6){$\gamma^\mu$}
\put(11.7,0){$p_1$} \put(13.7,0){$p_2$}% \put(14.8,-2){$(b)$}
\put(16,0){$p_3$} \put(18,0){$p_4$}
\end{picture}
\]
\caption{\textit{Feynman graphs}}
\label{f2}
\end{figure}

Several additional checks \cite{BFKZ,BK} have been performed to justify the
correspondences of operators and p-functions. The Feynman graph expansion of
the matrix element has been compared with the expansion of the exact result
given by the integral representation and always agreement has been found.
For example the 4-particle form factor of the sine Gordon $\varphi $-field
can be calculated in lowest order in $g$ by the Feynman graphs of figure \ref
{f2} using {Coleman's \cite{Co} formula} $\epsilon ^{\mu \nu }\partial _{\nu
}\varphi =-\frac{2\pi }{\beta }\overline{\psi }\gamma ^{\mu }\psi $. 
The result is 
\[
\langle \,0\,|\,\varphi \,(0)|\,p_{1},\dots ,p_{4}\,\rangle _{\bar{s}s\bar{s}%
s}^{in}=\frac{2\pi ig}{\beta }\frac{\sinh \frac{1}{2}\theta _{13}\sinh \frac{%
1}{2}\theta _{24}\sinh \frac{1}{2}(\theta _{12}+\theta _{34})}{%
\prod_{i<j}\cosh \frac{1}{2}\theta _{ij}}+O(g^{2}) 
\]
which agrees with the exact result. Another check is to calculate the
eigenvalues of charges corresponding to local operators. For example it can
be shown \cite{BFKZ} that with the p-functions above the higher charges
satisfy the eigenvalue equation 
\[
\left( \int dxJ_{L}^{0}(x)-\sum_{i=1}^{n}e^{L\theta _{i}}\right)
|\,p_{1},\dots ,p_{n}\rangle ^{in}=0 
\]
as is expected from the classical higher conservation laws.

\subsection*{Breather form factors}

It has been shown it the original article \cite{KW} that for particles which
do not possess backward scattering the generalized form factors can be
written as 
\begin{equation}
\mathcal{O}_{n}(\theta _{1},\dots ,\theta _{n})=K_{n}^{\mathcal{O}}(%
\underline{\theta })\prod_{1\leq i<j\leq n}F(\theta _{ij})  \label{2}
\end{equation}
where $F(\theta )$ is the two-particle form factor. It satisfies Watson's
equations 
\[
F(\theta )=F(-\theta )S(\theta )=F(2\pi i-\theta )
\]
where $S(\theta )$ is the diagonal two-particle S-matrix. For the lowest
breathers of the sine-Gordon model the S-matrix is given above and the
two-particle form factor is \cite{KW} 
\[
F_{bb}(\theta )=N\exp \int_{0}^{\infty }\frac{dt}{t}\,\frac{\cosh \frac{1}{2}%
t-\cosh (\frac{1}{2}+\nu )t}{\cosh \frac{1}{2}t\sinh t}\left( 1-\cosh
t\left( 1-\frac{\theta }{i\pi }\right) \right) 
\]
(normalized such that $F_{bb}(\infty )=1$). The K-function $K_{n}^{\mathcal{O%
}}(\underline{\theta })$ depends on the operator and it satisfies Watson's
equations for $S=1$. For simple cases the K-functions have been proposed in
the original article \cite{KW}. Smirnov \cite{Sm} used the bound state fusion
method to obtain a breather form factor formula. For the sinh-Gordon model
in several articles \cite{FMS,KM,MS,BL} K-functions where proposed for
various operators.

Starting with the general formula (\ref{1}) for soliton form factors and
using the bound state fusion method we derive \cite{BK} soliton-breather and
pure breather form factor formulae. The K-function for the case of the
lowest breathers turns out to be of the form 
\begin{equation}
K_{n}^{\mathcal{O}}(\underline{\theta })=\sum_{l_{1}=0}^{1}\dots
\sum_{l_{n}=0}^{1}(-1)^{l_{1}+\dots +l_{n}}\prod_{1\leq i<j\leq n}\left(
1+(l_{i}-l_{j})\frac{i\sin \pi \nu }{\sinh \theta _{ij}}\right) p_{n}^{%
\mathcal{O}}(\underline{\theta },\underline{l})  \label{3}
\end{equation}
where the breather p-function $\,p_{n}^{\mathcal{O}}(\underline{\theta },%
\underline{l})=\,p_{sol,2n}^{\mathcal{O}}(\underline{\tilde{\theta}},%
\underline{z})$ is obtained from the solitonic one with $\tilde{\theta}%
_{2i-1}=\theta _{i}+\tfrac{1}{2}i\pi (1-\nu )\,,\,\tilde{\theta}_{2i}=\theta
_{i}-\tfrac{1}{2}i\pi (1-\nu )a\,,z_{i}=\theta _{i}-\tfrac{1}{2}i\pi \left(
1-(-1)^{l_{1}}\nu \right) $. In this way we \cite{BK2} obtain the breather
p-functions for the {fundamental breather field} $\varphi (x)$, the {energy
momentum tensor} and {the $\infty $-many conserved currents} from the
solitonic p-functions above.

However, one may also assume a different point of view. Namely, one can
consider the form (\ref{3}) of the K-functions as an Ansatz and look for
breather p-functions such that the general form factor equations $(i)-(v)$
are satisfied. Doing this one will obtain a wider class of p-functions
corresponding to operators which are local with respect to the breather
field, but not necessarily local with respect to solitonic field. For
example we propose \cite{BK1}\footnote{%
For the sinh-Gordon model an analogous representation as (\ref{2}) together
with this p-function was obtained previously \cite{BL} by different methods.}
the breather p-function corresponding to the normal ordered exponentials of
the field $:\!e^{i\gamma \varphi }\!:\!(x)$ for generic real $\gamma $ 
\begin{equation}
:\!e^{i\gamma \varphi }\!:\,\leftrightarrow p_{n}^{(q)}(\underline{l}%
)=N_{n}^{(q)}\prod_{i=1}^{n}q^{(-1)^{l_{i}}}\;\text{with }q=\exp \left( i%
\frac{\pi \nu }{\beta }\gamma \right) \,.  \label{pq}
\end{equation}
Here and in the following $:\dots :$ denotes normal ordering with respect to
the physical vacuum which means in particular for the vacuum expectation
value $\langle \,0\,|\!:\!\exp i\gamma \varphi \!:\!(x)|\,0\,\rangle =1$. This
breather p-function is not related to a solitonic p-function of any local
operator (at least for generic $\gamma $). One easily calculates the
examples for $n=1,2$ explicitly 
\begin{eqnarray*}
\left[ e^{i\gamma \varphi }\right] _{1}(\theta ) &=&N_{1}^{(q)}\left(
q-1/q\right) \\
\left[ e^{i\gamma \varphi }\right] _{1}(\theta _{1},\theta _{2})
&=&N_{2}^{(q)}\left( q-1/q\right) ^{2}F_{bb}(\theta _{12})
\end{eqnarray*}
Expanding these relation in powers of $\gamma $ one obtains the p-functions
of all normal ordered powers of the field $\varphi $%
\begin{equation}
:\!\varphi ^{N}\!:\,\leftrightarrow p_{n}^{(N)}(\underline{l}%
)=N_{n}^{(N)}\left( \sum_{i=1}^{n}(-1)^{l_{i}}\right) ^{N}\;\text{with }%
N_{n}^{(N)}=\left( \frac{\pi \nu }{\beta }\right) ^{N}N_{n}^{(q)}  \label{pN}
\end{equation}
(since $N_{n}^{(q)}$ is independent on $q$ as we will see below). In
particular for $N=1$ and $n=1,2$ one calculates 
\begin{gather*}
\langle \,0\,|\,\varphi (0)\,|\,p_{1}\,\rangle =2N_{1}^{(1)} \\
\langle \,0\,|\,\varphi (0)\,|\,p_{1},p_{2}\,\,\rangle ^{in}=0\,.
\end{gather*}
Note that this breather p-function for $\varphi (x)$ is not the same as that
obtained from the solitonic one above using bound state fusion. However, it
can be proven \cite{BK2} that the form factors are the same in both cases. To
justify the proposal (\ref{pq}) and to calculate the normalization constants
we investigate the asymptotic behavior of form factors.

\subsection*{Asymptotic behavior of form factors for $\,:\!e^{i\gamma
\varphi }\!:$}

Let $\mathcal{O}=\,:\!\varphi ^{N}\!:$ be the normal ordered power of a
bosonic field. Set the rapidities as $\underline{\theta }=\lambda \theta
_{1}^{\prime },\dots ,\lambda \theta _{m}^{\prime },\theta _{1}^{\prime
\prime },\dots ,\theta _{n-m}^{\prime \prime }$ and let $\lambda \rightarrow
\infty $ then the asymptotic behavior of the n-boson form factor is 
\[
\left[ \varphi ^{N}\right] _{n}(\underline{\theta })=\sum_{K=0}^{N}\binom{N}{%
K}\left[ \varphi ^{K}\right] _{m}(\underline{\theta }^{\prime })\,\left[
\varphi ^{N-K}\right] _{n-m}(\underline{\theta }^{\prime \prime
})+O(e^{-\lambda }) 
\]
if the interaction is pure bosonic. This can be proven in any order of
perturbation theory as follows. The matrix element on the left hand side may
be written in terms of Feynman graphs as 
\[
\begin{array}{c}
{\unitlength3.6mm%
\begin{picture}(6,6) \put(3,2){\oval(6,2)} \put(3,5){\makebox(0,0){$\bullet$}} \put(3,5.8){\makebox(0,0){${\cal
O}=:\varphi^N:$}} \put(1,3){\line(1,1){2}} \put(2,3){\line(1,2){1}} \put(5,3){\line(-1,1){2}} \put(3,.5){$\dots$} \put(2.7,3.5){$N$} \put(1,0){\line(0,1){1}} \put(2,0){\line(0,1){1}} \put(5,0){\line(0,1){1}} \put(-2,0){$\theta_1+\lambda$} \put(5.4,0){$\theta_r$} \end{picture} 
}
\end{array}
~=\sum_{K=0}^{N}\binom{N}{K}~ 
\begin{array}{c}
\unitlength3.6mm\begin{picture}(11,6) \put(2,2){\oval(4,2)}
\put(8,2){\oval(4,2)} \put(5,5){\makebox(0,0){$\bullet$}}
\put(5,5.8){\makebox(0,0){${\cal O}=:\varphi^N:$}} \put(1,3){\line(2,1){4}}
\put(3,3){\line(1,1){2}} \put(7,3){\line(-1,1){2}} \put(9,3){\line(-2,1){4}}
\put(1.5,.5){$\dots$} \put(7.5,.5){$\dots$} \put(1.7,4){$K$}
\put(8,4){$N-K$} \put(1,0){\line(0,1){1}} \put(3,0){\line(0,1){1}}
\put(7,0){\line(0,1){1}} \put(9,0){\line(0,1){1}} \put(-.1,0){$\theta_1$}
\put(3.4,0){$\theta_s$} \put(9.4,0){$\theta_r$} \end{picture}
\end{array}
+\dots 
\]
where all other graphs not drawn have lines which connect both parts.
Weinbergs power counting theorem for bosonic Feynman graphs implies that
these contributions decrease for $\lambda \rightarrow \infty $ as $O(\lambda
^{k}e^{-\lambda })\,$. This behavior is also assumed to hold for the exact
form factors (the fact is that the `logarithmic terms' $\lambda ^{k}$ do not
show up for the exact expressions since the K-functions are meromorphic in
the $e^{\theta _{i}}$). Therefore for the exponentials of the boson field $%
:e^{i\gamma \varphi }:$ we have the asymptotic behavior 
\[
\left[ e^{i\gamma \varphi }\right] _{n}(\underline{\theta })=\left[
e^{i\gamma \varphi }\right] _{m}(\underline{\theta }^{\prime })\,\left[
e^{i\gamma \varphi }\right] _{n-m}(\underline{\theta }^{\prime \prime
})+O(e^{-\lambda })\,. 
\]
It is easy to see \cite{BK1,BK2} that our proposal (\ref{pq}) together with (%
\ref{2}) and (\ref{3}) satisfies this asymptotic behavior.\footnote{%
This type of arguments has been also used before \cite{KW,FMS,KM,MS}.} The
asymptotic behavior of other form factors is more complicated \cite{BK} in
particular if fermions are involved.

\subsection*{Normalization of form factors}

The normalization constants are obtained in the various cases by the
following observations:

\begin{enumerate}
\item[a)]  The recursion relation $(iii)$ relates $N_{n+2}$ and $N_{n}.$ For
a typical p-function as that for the exponentials of the field this means 
\begin{equation}
N_{n}^{(q)}=N_{n-2}^{(q)}\frac{2}{\sin \pi \nu F_{bb}(i\pi )}\quad (n\geq 3).
\label{N}
\end{equation}

\item[b)]  For a field annihilating a one-particle state the normalization
is given by the vacuum one-particle matrix element, in particular for the
fundamental breather field one has 
\[
\langle \,0\,|\,\varphi (0)\,|\,p\,\rangle =\sqrt{Z^{\varphi }} 
\]
where $Z^{\varphi }$ is the finite wave function renormalization constant.
For the sine-Gordon field it has been calculated in the original article \cite
{KW} 
\[
Z^{\varphi }=(1+\nu )\frac{\frac{\pi }{2}\nu }{\sin \frac{\pi }{2}\nu }\exp
\left( -\frac{1}{\pi }\int_{0}^{\pi \nu }\frac{t}{\sin t}dt\right) . 
\]

\item[c)]  If a local operator is related to an observable like a charge $%
Q=\int dx\,\mathcal{O}(x)$ we use the relation 
\[
\langle \,p^{\prime }\,|\,Q\,|\,p\,\rangle =q\langle \,p^{\prime
}\,|\,\,p\,\rangle . 
\]
This may be applied for example to the higher conserved charges.

\item[d)]  We use Weinberg's power counting theorem for bosonic Feynman
graphs. As discussed above this yields in particular the asymptotic behavior
for the exponentials of the boson field $\mathcal{O}=\,:\!e^{i\gamma \varphi
}\!:$ 
\[
\mathcal{O}_{n}(\theta _{1,}\theta _{2,}\dots )=\mathcal{O}_{1}(\theta
_{1})\,\mathcal{O}_{n-1}(\theta _{2,}\dots )+O(e^{-\func{Re}\theta _{1}}) 
\]
as $\func{Re}\theta _{1}\rightarrow \infty $ in any order of perturbation
theory. This behavior is also assumed to hold for the exact form factors.
Applying this formula iteratively we obtain from (\ref{3}) relations for the
normalization constants.
\end{enumerate}

\noindent \textbf{Examples: }

\begin{enumerate}
\item  For the normalization factors for exponentials of the field given by (%
\ref{pq}) one uses a) and d) \cite{BK1,BK2} 
\[
\left. 
\begin{array}{c}
N_{n}^{(q)}=N_{n-2}^{(q)}\dfrac{2}{\sin \pi \nu F_{bb}(i\pi )} \\ 
N_{n}^{(q)}=N_{1}^{(q)}N_{n-1}^{(q)}
\end{array}
\right\} \;\Rightarrow \;N_{n}^{(q)}=\left( \sqrt{Z^{\varphi }}\frac{\beta }{%
2\pi \nu }\right) ^{n} 
\]
where the identity $Z^{\varphi }F(i\pi )\beta ^{2}\sin \pi \nu =8\left( \pi
\nu \right) ^{2}$ has been used. Note that $N_{n}^{(q)}$is independent of $q$%
.

\item  For the normalization factors for the field given by (\ref{pN}) (for $%
N=1)$ one uses b) which implies 
\[
\langle \,0\,|\,\varphi (0)\,|\,p\,\rangle =2N_{1}^{(1)}=\sqrt{Z^{\varphi }}%
. 
\]
This is consistent with our proposals (\ref{pq}) and (\ref{pN}) and
justifies the identification $q=\exp \left( i\frac{\pi \nu }{\beta }\gamma
\right) $.
\end{enumerate}

\section{Some Results}

\subsection*{The quantum sine-Gordon equation}

We start with the local operator $:\!\sin \gamma \varphi \!:(x)=\frac{1}{2i}%
:\!\left( e^{i\gamma \varphi }-e^{-i\gamma \varphi }\right) \!:(x)$. For the
exceptional value $\gamma =\beta $ we find \cite{BK,BK2} that also $\Box
^{-1}\!:\!\sin \beta \varphi \!:(x)$ is local. Moreover the quantum
sine-Gordon field equation 
\begin{equation}
\Box \varphi (x)+\frac{\alpha }{\beta }:\!\sin \beta \varphi \!:(x)=0
\label{e}
\end{equation}
holds for all matrix elements, if the ``bare'' mass $\sqrt{\alpha }$ is
related to the renormalized mass by\footnote{%
Before such formula was found \cite{Fa,Za1} by different methods.} 
\begin{equation}
\alpha =m^{2}\frac{\pi \nu }{\sin \pi \nu }  \label{mass}
\end{equation}
where $m$ is the physical mass of the fundamental boson.

This is a sketch of the proof \cite{BK2} which uses induction and Liouville's
theorem. Consider the K-functions of the left hand side of (\ref{e}) 
\[
f_{n}(\underline{\theta })=-\sum e^{\theta _{i}}\sum e^{-\theta
_{i}}K_{n}^{(1)}(\underline{\theta })+\frac{\pi \nu }{\beta \sin \pi \nu }%
\frac{1}{2i}\left( K_{n}^{(q)}(\underline{\theta })-K_{n}^{(1/q)}(\underline{%
\theta })\right) \,. 
\]
The results of the previous section imply $f_{1}(\theta )=f_{2}(\underline{%
\theta })=0$ for $q=e^{i\pi \nu }$. As induction assumption we take $f_{n-2}(%
\underline{\theta }^{\prime \prime })=0$. The function $f_{n}(\underline{%
\theta })$ is meromorphic in terms of the $x_{i}=e^{\theta _{i}}$ with at
most simple poles at $x_{i}=\pm x_{j}$ since $\sinh \theta
_{ij}=(x_{i}+x_{j})(x_{i}-x_{j})/(2x_{i}x_{j})$. The residues of the poles
at $x_{i}=x_{j}$ vanish because of the symmetry under the exchange of $%
x_{i}\leftrightarrow x_{j}$. The residues at $x_{i}=-x_{j}$ are proportional
to $f_{n-2}(\underline{\theta }^{\prime \prime })$ because of the recursion
relation $(iii)$. Furthermore it can be shown \cite{BK2} that $f_{n}(%
\underline{\theta })\rightarrow 0$ for $x_{i}\rightarrow \infty $. Therefore 
$f_{n}(\underline{\theta })$ vanishes identically by Liouville's theorem.

The factor $\frac{\pi \nu }{\sin \pi \nu }$ in (\ref{mass}) modifies the
classical equation and has to be considered as a quantum correction. For the
sinh-Gordon model an analogous quantum field equation has been obtained
previously \cite{MS}\footnote{%
It should be obtained from (\ref{e}) by the replacement $\beta \rightarrow
ig $. However the relation between the bare and the renormalized mass
differs from the analytic continuation of (\ref{mass}) by a factor.}. Note
that in particular at the `free fermion point' $\nu \rightarrow 1~(\beta
^{2}\rightarrow 4\pi )$ this factor diverges, a phenomenon which is to be
expected by short distance investigations \cite{ST}. For fixed bare mass
square $\alpha $ and $\nu \rightarrow 2,3,4,\dots $ the physical mass goes
to zero. These values of the coupling are known to be specific: 1) the Bethe
Ansatz vacuum in the language of the massive Thirring model shows phase
transitions \cite{Ko} and 2) the model at these points is related \cite
{K3,LeC,Sm1} to Baxters RSOS-models which correspond to minimal conformal
models with central charge $c=1-6/(\nu (\nu +1))$.

\subsection*{The trace of the energy momentum tensor}

As a further operator equation we find \cite{BK1,BK2} that the trace of the
energy momentum tensor satisfies 
\begin{equation}
T_{~\mu }^{\mu }(x)=-2\frac{\alpha }{\beta ^{2}}\left( 1-\frac{\beta ^{2}}{%
8\pi }\right) \left( :\!\cos \beta \varphi \!:(x)-1\right) .  \label{T}
\end{equation}
Again this operator equations is to be understood as equations of all its
matrix elements. The equation is modified compared to the classical one by a
quantum correction $(1-\beta ^{2}/8\pi )$. As a consequence of this fact the
model will be conformal invariant in the limit $\beta ^{2}\rightarrow 8\pi $
for fixed bare mass square $\alpha $. This is related to a
Berezinski-Kosterlitz-Thouless \cite{KS} phase transition.

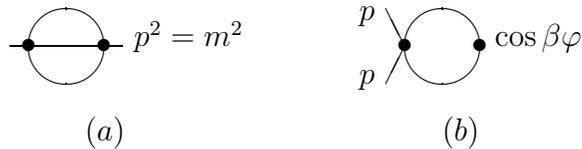
\begin{figure}[tbh]
\[
\begin{array}{l}
\unitlength5mm \begin{picture}(14,3) \put(1,2){\oval(2,2)}
\put(0,2){\makebox(0,0){$\bullet$}} \put(2,2){\makebox(0,0){$\bullet$}}
\put(-.5,2){\line(1,0){3}} \put(2.8,2){$p^2=m^2$} \put(1.5,-.5){$(a)$}
\put(11,2){\oval(2,2)} \put(10,2){\makebox(0,0){$\bullet$}}
\put(12,2){\makebox(0,0){$\bullet$}} \put(12.4,2){$\cos \beta \varphi$}
\put(10,2){\line(-1,2){.5}} \put(10,2){\line(-1,-2){.5}} \put(8.8,1){$p$}
\put(8.8,2.7){$p$} \put(11,-.5){$(b)$} \end{picture}
\end{array}
\]
\caption{Bosonic Feynman graphs}
\label{f8}
\end{figure}
All the results may be checked in perturbation theory by Feynman graph
expansions. In particular in lowest order the relation between the bare and
the renormalized mass (\ref{mass}) is given by figure \ref{f8} (a). It had
already been calculated in the original article \cite{KW}. The result is 
\[
m^{2}=\alpha \left( 1-\frac{1}{6}\left( \frac{\beta ^{2}}{8}\right)
^{2}+O(\beta ^{6})\right) 
\]
which agrees with the exact formula above. Similarly we check the quantum
corrections of the trace of the energy momentum tensor (\ref{T}) by
calculating the Feynman graph of figure \ref{f8} (b) with the result \cite{KW}
\[
\langle \,p\,|\,\!:\!\cos \beta \varphi \!:(0)-1|\,p\,\rangle =-\beta
^{2}\left( 1+\frac{\beta ^{2}}{8\pi }\right) +O(\beta ^{6}). 
\]
This again agrees with the exact formula above since the usual normalization
for the energy momentum given by c) implies $\langle \,p\,|\,T_{~\mu }^{\mu
}|\,p\,\rangle =2m^{2}$.

\section*{Acknowledgments}

We thank A.A. Belavin, J. Balog, V.A. Fateev,
R. Flume, A. Fring, R.H. Poghossian, F.A.
Smirnov, R. Schrader, B. Schroer and Al.B. Zamolodchikov for discussions.
One of authors (M.K.) thanks E. Seiler and P. Weisz for discussions and
hospitality at the Max-Planck Insitut f\"{u}r Physik (M\"{u}nchen), where
parts of this work have been performed. H.B. was supported by DFG,
Sonderforschungsbereich 288 `Differentialgeometrie und Quantenphysik' and
partially by grants INTAS 99-01459 and INTAS 00-561.


\begin{thebibliography}{99}
\bibitem{Heisen}  W. Heisenberg, \emph{Zeit. f\"{u}r Naturforschung} \textbf{%
1} (1946) 608.

\bibitem{Jost}  R. Jost, \emph{Helv. Phys. Acta} \textbf{20} (1947) 256.

\bibitem{Bargmann}  V. Bargmann, \emph{Rev. Mod. Phys.} \textbf{21} (1949)
488.

\bibitem{ELOP}  R.J. Eden, P.V. Landshoff, D.I. Olive and J.C. Polkinghorne, 
\emph{The analytic S-matrix} (CUP, Cambridge, 1966).

\bibitem{Barton}  G. Barton, \emph{Introduction to Dispersion Techniques in
Field Theory} (W.A. Benjamin Inc., New York, 1965).

\bibitem{K2}  M. Karowski, \emph{The bootstrap program for 1+1 dimensional
field theoretic models with soliton behavior}, in 'Field theoretic methods
in particle physics', ed. W. R\"{u}hl, (Plenum Pub. Co., New York ,1980).

\bibitem{KTTW}  M. Karowski, H.J. Thun, T.T. Truong and P. Weisz, \emph{%
Phys. Lett.} \textbf{B67} (1977) 321.

\bibitem{VG}  S. Vergeles and V. Gryanik, \emph{Sov. Journ. Nucl. Phys.} 
\textbf{23} (1976) 704.

\bibitem{W}  P. Weisz, \emph{Nucl. Phys.} \textbf{B122} (1977) 1.

\bibitem{KW}  M. Karowski and P. Weisz, \emph{Nucl. Phys.} \textbf{B139}
(1978) 445.

\bibitem{Sm}  F.A. Smirnov \emph{'Form Factors in Completely Integrable
Models of Quantum Field Theory', Adv. Series in Math. Phys.} \textbf{14},
World Scientific 1992.

\bibitem{BFKZ}  H. Babujian, A. Fring, M. Karowski and A. Zapletal, \emph{%
Nucl. Phys.} \textbf{B538} [FS] (1999) 535-586.

\bibitem{CM}  J.L. Cardy and G. Mussardo, \emph{Phys. Lett.} \textbf{B225}
(1989) 275. \emph{Nucl. Phys.} \textbf{B340} (1990) 387.

\bibitem{ReshSm}  N. Reshetikin and F. Smirnov, \emph{Commun. Math. Phys. } 
\textbf{131} (1990) 157.

\bibitem{Smirnov4}  F.A. Smirnov, \emph{Nucl. Phys.} \textbf{B337} (1990)
156, \emph{Int. J. Mod. Phys.} \textbf{A9} (1994) 5121, \emph{Nucl. Phys.} 
\textbf{B453} (1995) 807.

\bibitem{YLZam}  Al.B. Zamolodchikov, \emph{Nucl. Phys.} \textbf{B348}
(1991), 619.

\bibitem{YZ}  V.P. Yurov and Al. B. Zamolodchikov, \emph{Int. J. Mod. Phys.} 
\textbf{A6} (1991) 4557.

\bibitem{DeDe}  C. Destri and H.J. De Vega, \emph{Nucl. Phys.} \textbf{B358}
(1991) 251.

\bibitem{BB}  O. Babelon and D. Bernard, \emph{Phys. Lett.} \textbf{B288}
(1992) 113.

\bibitem{FMS}  A. Fring, G. Mussardo and P. Simonetti,  \emph{Nucl. Phys.}%
\textbf{\ B393} (1993) 413, \emph{Phys. Lett.} \textbf{B307} (1993) 83.

\bibitem{KM}  A. Koubek and G. Mussardo, \emph{Phys. Lett.} \textbf{B311}
(1993) 193.

\bibitem{Ba}  J. Balog, \emph{\ Phys. Lett.} \textbf{B300} (1993) 145.

\bibitem{Anni}  A. Koubek, \emph{Nucl. Phys.} \textbf{B428} (1994) 655.

\bibitem{MS}  G. Mussardo and P. Simonetti, \emph{Int. J. Mod. Phys.}\textbf{%
\ A9} (1994) 3307-3338

\bibitem{Ahn}  C. Ahn, \emph{Nucl. Phys.} \textbf{B422} (1994) 449.

\bibitem{BH}  J. Balog and T. Hauer, \emph{\ Phys. Lett.} \textbf{B337}
(1994) 115.

\bibitem{LE}  A. LeClair and C. Efthimiou, \emph{Commun. Math. Phys. } 
\textbf{171} (1995) 531.

\bibitem{DM}  G. Delfino and G. Mussardo, \emph{Nucl. Phys.} \textbf{B455}
(1995) 724.

\bibitem{AMV}  C. Acerbi, G. Mussardo and A. Valleriani, \emph{Int. J. Mod.
Phys.} \textbf{A11} (1996) 5327; \emph{J. Phys.} \textbf{A30} (1997) 2895.

\bibitem{LSS}  F. Lesage, H. Saleur and S. Skorik, \emph{Nucl. Phys.} 
\textbf{B474} (1996) 602.

\bibitem{DSC}  G. Delfino, P. Simonetti and J.L Cardy, \emph{\ Phys. Lett.} 
\textbf{B387} (1996) 327.

\bibitem{Konn}  H. Konno, \emph{Nucl. Phys.} \textbf{B432} (1995) 457; 
\newline
\emph{Degeneration of the Elliptic Algebra $A_{p,q}(\hat{sl}_{2})$ and Form
Factors in the sine Gordon Theory} hep-th/9701034.

\bibitem{OO}  T. Oota, \emph{Nucl. Phys.} \textbf{B466} (1996) 361.

\bibitem{BBS}  O. Babelon, D. Bernard and F.A. Smirnov, \emph{Nucl. Phys.
Proc. Suppl.} \textbf{58} (1997) 21.

\bibitem{MSm}  P. Mejean and F.A. Smirnov, \emph{Int. J. Mod. Phys.} \textbf{%
A12} (1997) 3383.

\bibitem{Smcl}  F.A. Smirnov, \emph{Quasi-classical Study of Form Factors in
finite Volume} hep-th/9802132.

\bibitem{Luk3}  S. Lukyanov, \emph{\ Mod. Phys. Lett.} \textbf{A12} (1997)
2543; \emph{\ Phys. Lett.} \textbf{B408} (1997) 192.

\bibitem{Ace}  C. Acerbi, \emph{Nucl. Phys.} \textbf{B497} (1997) 589.

\bibitem{Pil}  M. Pillin, \emph{Nucl. Phys.} \textbf{B497} (1997) 569, \emph{%
\ Lett. Math. Phys. } \textbf{43} (1998) 569.

\bibitem{Luk2}  V. Brazhnikov and S. Lukyanov, \emph{Nucl. Phys.} \textbf{%
B512} (1998) 616.

\bibitem{Qu}  Y. Quano, \emph{J. Phys.} \textbf{A31} (1998) 1791.

\bibitem{LZ}  S Lukyanov and A.B. Zamolodchikov, hep-th /0102079.

\bibitem{KLP}  S. Khoroshkin, D. Lebedev and S. Pakuliak, \emph{Lett. Math.
Phys. }\textbf{41} (1997) 31-47.

\bibitem{BL}  V. Brazhnikov and S. Lukyanov, $\emph{Nucl.Phys.}$\textbf{\
B512} (1998) 616-636.

\bibitem{NPT}  A. Nakayashiki, S. Pakuliak and V. Tarasov, \emph{Annales de
l'Institut Henri Poincar\'{e} }\textbf{71 }N4 (1999) 459-496.

\bibitem{NT}  A. Nakayashiki, Y. Takeyama, \emph{On form factors of the }$%
SU(2)$ \emph{invariant Thirring model }math-ph/0105040.

\bibitem{BK1}  H.M. Babujian and M. Karowski, Phys. Lett. \textbf{B 411}
(1999) 53-57.

\bibitem{BK}  H. Babujian and M. Karowski, \emph{Exact Form Factors in
Integrable Quantum Field Theories: the Sine-Gordon Model (II)}, Sfb 288 -
preprint 506, hep-th/0105178.

\bibitem{BK2}  H. Babujian and M. Karowski, \emph{Exact Form Factors in
Integrable Quantum Field Theories: the Sine-Gordon Model (III)}, in
preparation.

\bibitem{GNT}  A.O. Gogolin, A.A Nersesyan and A.M. Tsvelik, \emph{%
'Bosonization in Strongly Correlated Systems', }Cambridge University Press
(1999).

\bibitem{CET}  D. Controzzi, F.H.L. Essler and A.M. Tsvelik, \emph{%
'Dynamical Properties of one dimensional Mott Insulators', }%
cond-math/0011439.

\bibitem{B-J}  R.Z. Bariev, {\em Phys. Lett.} {\bf55A} (1976) 456;\newline
B. McCoy, C.A. Tracy and T.T. Wu, {\em Phys. Rev. Lett.} {\bf 38} (1977) 783;
\newline
M. Sato, T. Miva and M. Jimbo, {\em Proc. Japan Acad.} {\bf 53A} (1977) 6.

\bibitem{BKW}  B. Berg, M. Karowski and P. Weisz, \emph{Phys. Rev.} \textbf{%
D19} (1979) 2477.

\bibitem{Korepin}  V.E. Korepin and N.A. Slavnov, \emph{J. Phys.} \textbf{A31%
} (1998) 9283;\newline
V.E. Korepin and T.Oota, \emph{J. Phys.} \textbf{A31} (1998) L371;\newline
T. Oota, \emph{J. Phys.} \textbf{A31} (1998) 7611.

\bibitem{BA}  H. Bethe, \emph{Zeit. der Physik} \textbf{71 }(1931) 205;%
\newline
L. Faddeev, \emph{Sov. Sci. Reviews} \textbf{C1} (1980) 107; \newline
V.E. Korepin, N.M. Bogoliubov and A.G. Izergin \emph{Quantum Inverse
Scattering Method and Correlation Functions}, (CUP,Cambridge, 1993); \newline
L.D. Faddeev, \emph{How Algebraic Bethe Ansatz works for integrable model}
Les Houches lecture notes 1995 hep-th/9605187.

\bibitem{OSBA}  H.M. Babujian, \emph{Correlation functions in WZNW model as
a Bethe wave function for the Gaudin magnets}, in: Proc. XXIV Int. Symp.
Ahrenshoop, Zeuthen 1990;\newline
H.M. Babujian, \emph{J. Phys.} \textbf{A26} (1993) 6981; H.M. Babujian and
R. Flume, \emph{Mod. Phys. Lett.} \textbf{A9} (1994) 2029.\newline
N. Reshetikin, \emph{\ Lett. Math. Phys. } \textbf{26} (1992) 153.

\bibitem{BKZ}  H. Babujian, M. Karowski and A. Zapletal, \emph{J. Phys.} 
\textbf{A30} (1997) 6425.

\bibitem{Co}  S. Coleman, \emph{Phys. Rev.} {\bf D11} (1975) 2088.

\bibitem{Za}  A.B. Zamolodchikov, \emph{JETP Lett.} \textbf{25} (1977) 468.

\bibitem{K1}  M. Karowski, \emph{\ Nucl. Phys.} \textbf{B153} (1979).

\bibitem{DHNKF}  R. Dashen, B. Hasslacher and A. Neveu, \emph{Phys. Rev.} 
\textbf{D10} (1974) 4114, 4130,4138; \textbf{D11} (1975) 3424;\newline
V.E. Korepin and L.D. Faddeev, \emph{Theor. Math. Phys.} \textbf{25} (1975)
1039.

\bibitem{KT}  M. Karowski and H.J. Thun, \emph{Nucl. Phys.} \textbf{B130}
(1977) 295.

\bibitem{LSZ}  H. Lehmann, K. Symanzik and W. Zimmermann, {\em Nuovo Cimento} 
\textbf{1} (1955) 205; \textbf{6} (1957) 319.

\bibitem{Las}  M.Yu. Lashkevich, \emph{Sectors of Mutually Local Fields in
Integrable Models of Quantum Field Theory, }LANDAU-94-TMP-4, hep-th/9406118.

\bibitem{Q}  T. Quella, \emph{Formfaktoren und Lokalit\"{a}t in integrablen
Modellen der Quan\-ten\-feld\-theo\-rie in 1+1 Dimensionen,} Diploma thesis
FU-Berlin (1999) unpublished.

\bibitem{Wa}  K.M. Watson, \emph{Phys. Rev.} \textbf{95} (1954) 228.

\bibitem{Fa}  V.A.Fateev, {\em Phys. Lett.} {\bf B 324} (1994) 45-51.

\bibitem{Za1}  Al.B. Zamolodchikov, \emph{Int. Journ. of Mod. Phys.}\textbf{%
\ A10} (1995) 1125-1150.

\bibitem{ST}  B. Schroer and T. Truong, \emph{Phys. Rev.} \textbf{15} (1977)
1684.

\bibitem{Ko}  V. E. Korepin, \emph{Commun. Math. Phys.} \textbf{76} (1980)
165.

\bibitem{K3}  M. Karowski,\emph{\ Nucl. Phys.} \textbf{B300} [FS22] (1988)
473; \newline
---, \emph{Yang-Baxter algebra - Bethe ansatz - conformal quantum field
theories - quantum groups}, in `Quantum Groups', Lecture Notes in Physics,
Springer (1990) p. 183.

\bibitem{LeC}  A. LeClair, \emph{Phys. Lett. }\textbf{B230} (1989) 103-107.

\bibitem{Sm1}  F.A. Smirnov, \emph{Commun. Math. Phys. }\textbf{131} (1990)
157-178.

\bibitem{KS}  J.M. Kosterlitz and J.P. Thouless, \emph{Journ. Phys. }\textbf{%
C6} (1973) 118.
\end{thebibliography}
\end{document}